\documentclass[prd,aps,eqsecnum,floatfix,nofootinbib,preprint,tightenlines]{revtex4}

\usepackage{graphicx}
\usepackage{amsmath}

\def\bibi{\bibitem}

\def\a{\alpha}

\def\c{\chi}
\def\d{\delta}

\def\m{\mu}
\def\n{\nu}

\def\p{\pi}                     
\def\r{\rho}                    
\def\t{\tau}

\def\P{\Pi}

\def\cbo{{\,\raise-.15ex\Sc [\,}}                       

\def\ddt#1{{\buildrel {\hbox{\LARGE .\kern-2pt.}} \over {#1}}}

\def\ie{\mbox{\it i.e.}}

\def\Im{{\rm Im\,}}

\def\floatcaption#1#2{ \caption{ #2 \ [#1] \label{#1}} }
\def\floatcaption#1#2{ \caption{#2 \label{#1}} }

\def\bibi{\bibitem}    

\def\ttl#1{{\it #1}}
\def\ttl#1{}

\long\def\symbolfootnote[#1]#2{\begingroup%
\def\thefootnote{\fnsymbol{footnote}}\footnote[#1]{#2}\endgroup}

\long \def \blockcomment #1\endcomment{}

\def\seef{{\it cf.}}

\begin{document}

\begin{center}
\begin{boldmath}
{\large\bf Model-independent parametrization of the hadronic
vacuum polarization and $g-2$ for the muon on the lattice}\\[0.4cm]
\end{boldmath}
\vspace{3ex}
{Christopher~Aubin,$^a$ Thomas~Blum,$^b$
Maarten~Golterman,$^c$  Santiago~Peris$^c$%
\symbolfootnote[2]{Permanent address: Department of Physics, Universitat Aut\`onoma de Barcelona, E-08193 Bellaterra, Barcelona, Spain}
\\[0.1cm]
{\it
\null$^a$Department of Physics\\ Fordham University, Bronx,
New York, NY 10458, USA\\
\null$^b$Physics Department\\
University of Connecticut, Storrs, CT 06269, USA\\
\null$^c$Department of Physics and Astronomy\\
San Francisco State University, San Francisco, CA 94132, USA}}
\\[6mm]
{ABSTRACT}
\\[2mm]
\end{center}
\begin{quotation} 
The leading hadronic contribution to the muon anomalous magnetic moment is given by a weighted integral over euclidean momentum of the hadronic vacuum polarization.   This integral is dominated by
momenta of order the muon mass.   
Since the finite volume in lattice QCD makes it difficult to compute the vacuum polarization at a large number of low momenta with high statistics (combined with the fact that one cannot compute it at zero momentum), a parametrization of the vacuum polarization is required to extrapolate the data.
A much used functional form  is based on vector
meson dominance, which introduces model dependence into the lattice
computation of the magnetic moment.   Here we introduce a model-independent extrapolation method, and present a few first tests of this new
method.

\end{quotation}

\vfill
\eject
\setcounter{footnote}{0}

\section{\label{introduction} Introduction}
Recently, there has been an increased interest in the lattice computation of the hadronic contributions to the anomalous magnetic moment $a_\m$ of the muon
\cite{TB2003,AB2007,QCDSF2004,FJPR2011,BDKZ2011,DJJW2012}.   The aim is to 
provide a first-principles computation of the hadronic contributions to $a_\m$
from lattice QCD assuming the Standard Model, with full control of the error.   Since the experimental
value of $a_\m$ is known with great accuracy \cite{BNL}, this would provide an
interesting test of the Standard Model if the theoretical computation can be
carried out with a comparable precision \cite{review}.

The dominant hadronic contribution comes from the hadronic vacuum polarization, and can be written as \cite{ER,TB2003}\footnote{Our sign convention for $\P(Q^2)$ is opposite to that of Ref.~\cite{AB2007}.}
\begin{subequations}
\begin{eqnarray}
\label{amu}
a_\m^{\rm HLO}&=&4\a^2\int_0^\infty dQ^2\,f(Q^2)\left(\P(0)-\P(Q^2)\right)\ ,\label{amua}\\
f(Q^2)&=&\frac{m_\m^2Q^2Z^3(Q^2)(1-Q^2Z(Q^2))}{1+m_\m^2Q^2Z^2(Q^2)}\ ,\label{amub}\\
Z(Q^2)&=&\frac{\sqrt{Q^4+4m_\m^2Q^2}-Q^2}{2m_\m^2Q^2}\ ,\label{amuc}\\
\P(Q^2)&=&\sum_{i=u,d,s}q_i^2\P_i(Q^2)\ ,\label{amud}
\end{eqnarray}
\end{subequations}
where $\P_i(Q^2)$, defined by
\begin{equation}
\label{vacpol}
\P_{\m\n;i}(Q)=(Q^2\d_{\m\n}-Q_\m Q_\n)\P_i(Q^2)\ ,
\end{equation}
is the (flavor-diagonal) vacuum polarization for quark flavor $i$.  In Eq.~(\ref{amud}), $q_i$ is the electric charge of quark $i$ in units of the 
electron charge, $\a$ is the fine-structure constant, and $m_\m$ is the 
muon mass.   Disconnected contributions are expected to be small
\cite{DJ2010} and have been neglected in Eq.~(\ref{amud}).\footnote{The methods
developed in this article also apply if disconnected parts are included as well.
For numerical
estimates of disconnected contributions, see Ref.~\cite{FJPR2011}.}   
Of course, Eq.~(\ref{amu}) also applies
to the electron and tau magnetic moments, if one replaces the muon mass
by the electron or tau mass.   

In principle, the integral in Eq.~(\ref{amua}) could be approximated by a sum, if  lattice values of $\P(Q^2)$ at sufficiently many low values of $Q^2$ (of order $m_\m^2$,
the region that dominates the integral) could be computed with high precision.
At present, this is not yet possible.  Instead, all lattice computations of 
$a_\m^{\rm HLO}$ rely on fitting the lattice data to a functional form for
$\P(Q^2)$, which is then used to compute the integral.   The most successful approach to date has been to choose a
functional form  inspired by vector-meson dominance (VMD), with either
only a contribution from the $\r$ pole (possibly dressed up with pion loop
contributions \cite{AB2007}), or with contributions from the $\r$ and the $\r'$
\cite{BDKZ2011}, where the
$\r$ mass is fixed to its value as independently determined on the same ensemble of gauge-field configurations.

However, this introduces a model-dependent element in what is supposed to 
be a first-principles computation in lattice QCD.   This results in a systematic
error afflicting lattice computations of $a_\m^{\rm HLO}$ which is difficult to
quantify.\footnote{As already discussed in Ref.~\cite{AB2007}, chiral
perturbation theory is of little help in this case.}   While VMD fits to $\P(Q^2)$ 
look very good, and lead to quite small statistical errors, $a_\m^{\rm HLO}$ is extremely sensitive to the behavior of
the fitted $\P(Q^2)$ at very small $Q^2$, and thus to any systematics affecting
the small-$Q^2$ behavior of $\P(Q^2)$.
Therefore, it would be very
nice if a functional form of $\P(Q^2)$ could be constructed that is solely based on known mathematical properties of the vacuum polarization, and which can be systematically improved if data with higher precision become available.

It turns out that such a method exists.   It is based on the well-known observation that the vacuum polarization can be expressed in terms of a positive spectral function through a (once-subtracted) dispersion relation.   This  makes it possible to express the
vacuum polarization $\P(Q^2)$ in a form for which a {\it convergent} sequence
of Pad\'e approximants (PAs) is known to exist.   Moreover, the convergence is uniform
for any compact region in the complex plane excluding the cut along the negative real axis.   This includes, in particular, any finite interval in euclidean $Q^2$ between $0$ and $\infty$.   Since the contribution from say the region $Q^2\ge 3$~GeV$^2$ to $a_\m$ is much smaller than currently attainable errors, this
is sufficient to employ this observation for the computation of $a_\m$. 
Our goal is to present an exploration of this observation,
using examples of available data for $\P(Q^2)$.

This article is organized as follows.   In Sec.~\ref{PA} we review elements of the necessary
mathematical theory, beginning with the observation that $\P(Q^2)$ can be
written in terms of a Stieltjes function, for which a converging sequence of
PAs is known to exist.
In Sec.~\ref{strategy} we explain how we will apply
this construction in order to carry out fits to numerical data for $\P(Q^2)$.
Section~\ref{tests} reports on two examples of such fits.   We discuss future
prospects of this approach in our concluding section.

\section{\label{PA} Stieltjes functions and Pad\'e approximants}
In this section, we review the necessary elements of the theory of
Pad\'e approximants (PAs) for functions that can be written as a
Stieltjes integral with a finite radius of convergence.   A good review is 
Ref.~\cite{BG1996}; for multi-point PAs we refer to Refs.~\cite{GB1969,MB1973}.

\subsection{\label{Stieltjes} Stieltjes functions}
Consider the function
\begin{equation}
\label{stieltjes}
\Phi(z)=\int_0^{1/R} \,\frac{d\n(\t)}{1+\t z}\ ,
\end{equation}
with $\n(\t)$ some real, bounded, non-decreasing function
on the interval $[0,1/R]$, taking infinitely many values on that interval.   The function $\Phi(z)$ then
is a Stieltjes function, and it is analytic everywhere in the complex plane
except on the negative real axis for $z\le -R$.   
The function $\Phi(z=Q^2)$ decreases monotonically as a function of $Q^2$
for $Q^2\in(-R,\infty)$. 

The vacuum polarization $\P(Q^2)$ can be expressed in terms of $\Phi$
through a once-subtracted dispersion relation
\begin{eqnarray}
\label{disprel}
\P(Q^2)&=&\P(0)-Q^2\Phi(Q^2)\ ,\\
\Phi(Q^2)&=&\int_{4m_\p^2}^\infty dt\,\frac{\r(t)}{t(t+Q^2)}\ ,\nonumber
\end{eqnarray}
where $\r(t)$ is the spectral function, which, of course, satisfies the
constraint $\r(t)\ge 0$ for $4m_\p^2\le t<\infty$.
This can be seen by changing variables $\t=1/t$ in the integral, taking
$R=4m_\p^2$, and choosing
\begin{eqnarray}
\label{rho}
d\n(\t)&=&\r(1/\t)\,d\t\ ,\\
\r(t)&=&\frac{1}{\p}\,\Im \P(t)\ .\nonumber
\end{eqnarray}

Let us consider an ordered sequence of positive values $Q^2_i$ of the
variable $Q^2$, with $i\in\{1,\dots,P\}$
and $0\le Q^2_1<Q^2_2<\dots<Q^2_P$, and assume that the function
$\Phi(Q^2)$ is known at these points.   We may now construct a sequence of
Stieltjes functions as follows.   We begin by defining a function 
$\Psi_1(Q^2)$ by writing
$\Phi(Q^2)$ as
\begin{equation}
\label{Psi1}
\Phi(Q^2)=\frac{\Phi(Q^2_1)}{1+(Q^2-Q_1^2)\Psi_1(Q^2)}\ .
\end{equation}
Then $\Psi_1(Q^2)$ is also a Stieltjes function \cite{GB1969}.   Moreover,
$\Psi_1(Q^2)$ is positive on the interval $[-R,\infty)$, and, on that interval,
has an upper bound
\begin{equation}
\label{ub}
\Psi_1(Q^2)\le\Psi_1(-R)\le\frac{1}{R+Q_1^2}\ ,\qquad Q^2\in[-R,\infty)\ .
\end{equation}
This follows from the requirement that $\Phi(Q^2)$ not have a singularity
on the real axis for $Q^2>-R$, which implies that
\begin{equation}
\label{limit}
\lim_{Q^2\downarrow-R}(Q_1^2-Q^2)\Psi_1(Q^2)\le 1\ .
\end{equation}
Clearly, a sequence of Stieltjes functions $\Psi_i(Q^2)$, $i\in\{1,\dots,P\}$,
can be constructed by iteration:
\begin{equation}
\label{iter}
\Psi_{i-1}(Q^2)=\frac{\Psi_{i-1}(Q^2_i)}{1+(Q^2-Q^2_i)\Psi_i(Q^2)}\ ,
\qquad i\in\{2,\dots,P\}\ ,
\end{equation}
which on the interval $[-R,\infty)$ satisfy
\begin{equation}
\label{ineq}
0\le\Psi_i(Q^2)\le\Psi_i(-R)=\frac{1}{R+Q_i^2}\left(1-\frac{\Psi_{i-1}(Q_i^2)}{\Psi_{i-1}(-R)}\right)\ ,
\quad i\in\{2,\dots,P\}\ ,
\end{equation}
where $\Psi_{i-1}(-R)=\lim_{Q^2\downarrow -R}\Psi_{i-1}(Q^2)$.

Equation~\ref{Psi1} defines $\Psi_1(Q^2_2)$ in terms of $\Phi(Q_2^2)$
and $\Phi(Q_1^2)$.   Likewise, in general, $\Psi_{i-1}(Q_i^2)$ can
be expressed in terms of the values $\Phi(Q_j^2)$, $j\in\{1,\dots,i\}$
by using Eq.~(\ref{iter}) recursively.

Applying Eq.~(\ref{iter}), the original function $\Phi(Q^2)$ can be written as a
continued fraction
\begin{equation}
\label{frac}
\Phi(Q^2)=\frac{\Phi(Q_1^2)}{1+\mbox{\Large{${\frac{(Q^2-Q_1^2)\Psi_1(Q_2^2)}
{1+\ { }_{{\ddots}_{{\ \frac{(Q^2-Q_{P-1}^2)\Psi_{P-1}(Q_P^2)}{1+(Q^2-Q_P^2)\Psi_P(Q^2)}}}}}}$}}}\ .
\end{equation}
As already observed above, $\Psi_1(Q_2^2),\dots,\Psi_{P-1}(Q_P^2)$
can be expressed in terms of the values of the function $\Phi(Q^2)$
at the points $Q_2^2,\dots,Q_P^2$.

\subsection{\label{multipoint} Multi-point Pad\'e's}
A rational (or Pad\'e) approximation to the function $\Phi(Q^2)$ can be constructed by
setting $\Psi_P(Q^2)$
in Eq.~(\ref{frac}) equal to its lower bound (\ie, zero), or its upper bound,
given by Eq.~(\ref{ineq}).   A rational approximation $R^N_M(Q^2)$ is the ratio of two polynomials of degrees $N$ and $M$
\begin{equation}
\label{defR}
R^N_M(Q^2)=\frac{\sum_{n=0}^N a_nQ^{2n}}{\sum_{n=0}^{M-1} b_nQ^{2n}
+Q^{2M}}\ .
\end{equation}
We will refer to $R^N_M(Q^2)$ as an $[N,M]$ PA.\footnote{Redundancy between the coefficients $a_n$ and $b_n$ is removed by choosing 
one of them equal to $1$.   Here we choose $b_{M}=1$.}

If we choose $\Psi_i(Q^2)=0$, the expression in Eq.~(\ref{iter}) yields a $[0,0]$
PA for $\Psi_{i-1}(Q^2)$.   Working
back to the original function, this choice leads to a PA for the function
$\Phi(Q^2)$.   If the number of points $Q_i^2$, $i\in\{1,\dots,P\}$ is even,
$P=2k$, starting with $\Psi_P(Q^2)=0$ yields a $[k-1,k]$ PA.  Indeed,
for a $[k-1,k]$ PA we need to solve for $k$ coefficients $a_n$ and $k$
coefficients $b_n$ in Eq.~(\ref{defR}), for a total of $P=2k$, determined by the
values $\Phi(Q_i^2)$, $i\in\{1,\dots,P\}$.  Likewise, for $P=2k+1$ odd, the
procedure yields a $[k,k]$ PA.   In short, from $\Psi_P(Q^2)=0$ one obtains
a $[\lfloor(P-1)/2\rfloor,\lfloor P/2\rfloor]$ PA, where $\lfloor x\rfloor$ denotes the integer part of $x$.

These ``standard''
multi-point PAs were studied in
Refs.~\cite{GB1969,MB1973}.   By construction,
they are exact at the values $Q^2=Q_i^2$, \ie,
the PA takes precisely the values $\Phi(Q_i^2)$ at these values of $Q^2$.
Moreover, these PAs converge to the function $\Phi(Q^2)$.   More precisely, if we consider
a sequence of standard multi-point PAs constructed from the values of $\Phi(Q^2)$ at a collection of points $Q_1^2<Q_2^2<\dots<Q_P^2<Q_*^2<\infty$
with $\lim_{P\to\infty} Q_P^2=Q_*^2$, the PAs converge uniformly on any closed and bounded domain in the complex $Q^2$-plane excluding the cut
$-\infty<Q^2\le-R$ \cite{MB1973}, for $P\to\infty$.

If we choose $\Psi_P(Q^2)$ equal to the upper bound of Eq.~(\ref{ineq}), the expression in Eq.~(\ref{iter}) yields a $[0,1]$
PA for $\Psi_{P-1}(Q^2)$.   Again working
back to the original function, this choice also leads to a PA for the function
$\Phi(Q^2)$.  Now if the number of values $Q_i^2$, $i\in\{1,\dots,P\}$ is even,
$P=2k$, this yields a $[k,k]$ PA.  The counting
argument is analogous to that above, but now this PA has, by construction,
a pole at $Q^2=-R$, which provides the extra information needed
to find the $2k+1$ coefficients in Eq.~(\ref{defR}) for this case.
Likewise, for $P=2k+1$ odd, the procedure yields a $[k,k+1]$ PA.
In short, in this case we obtain a $[\lfloor P/2\rfloor,\lfloor(P+1)/2\rfloor]$ PA.  
These ``complementary'' multi-point PAs are also exact at the values $Q^2=Q_i^2$; they were introduced in Ref.~\cite{MB1973}.

If, given $P$ values $\Phi(Q_i^2)$, $i\in\{1,\dots,P\}$, the standard PA is
written as 
\begin{equation}
\label{standard}
R^{\lfloor(P-1)/2\rfloor}_{\lfloor P/2\rfloor}(Q^2)=\frac{A_P(Q^2)}{B_P(Q^2)}\ ,
\end{equation}
defining the polynomials $A_P(Q^2)$ and $B_P(Q^2)$, the complementary
PA can be written as \cite{MB1973}
\begin{equation}
\label{compl}
R^{\lfloor P/2\rfloor}_{\lfloor(P+1)/2\rfloor}(Q^2)=\frac{(R+Q_P^2)B_{P-1}(-R)A_P(Q^2)
+(Q^2-Q_P^2)B_P(-R)A_{P-1}(Q^2)}{(R+Q_P^2)B_{P-1}(-R)B_P(Q^2)
+(Q^2-Q_P^2)B_P(-R)B_{P-1}(Q^2)}\ .
\end{equation}
As already mentioned, both these PAs are exact at the points $Q_i^2$.
Moreover, the complementary PAs have a pole at $Q^2=-R$, as can be
seen from Eq.~(\ref{compl}).  Between the points $Q_i^2$ and $Q_{i+1}^2$,
as well as between $-R$ and $Q_1^2$ and between $Q_P^2$ and $\infty$,
the standard and complementary PAs provide an upper and lower bound
to the original function $\Phi(Q^2)$.\footnote{These bounds are optimal \cite{GB1969,MB1973}.}
Which is the lower bound and which the upper bound alternates as one
progresses through these $P+1$ intervals from $-R$ to $\infty$ \cite{GB1969,MB1973}.

\subsection{\label{parametrization} Parametrization}
All the poles of our standard PAs should have their poles on the negative
real axis, at locations $Q^2\le -R$.   Indeed, one can prove \cite{BG1996,MB1973} that these PAs
can be written in the form
\begin{equation}
\label{standpar}
R^{\lfloor(P-1)/2\rfloor}_{\lfloor P/2\rfloor}(Q^2)=a_0+\sum_{n=1}^{\lfloor P/2\rfloor}\frac{a_n}{b_n+Q^2}\ ,
\end{equation}
with $a_0=0$ for $P$ even, and
\begin{eqnarray}
\label{cond}
&&a_n>0\ , \qquad n\in\{1,\dots,\lfloor P/2\rfloor\}\ ,\\
&&b_{\lfloor P/2\rfloor}>b_{\lfloor P/2\rfloor-1}>\dots>b_1\ge R\ .\nonumber
\end{eqnarray}
Once these parameters have been obtained for the $[\lfloor(P-2)/2\rfloor,\lfloor(P-1)/2\rfloor]$
and $[\lfloor(P-1)/2\rfloor,\lfloor P/2\rfloor]$ PAs, the complementary $[\lfloor P/2\rfloor,\lfloor(P+1)/2\rfloor]$ PA
can be obtained from Eq.~(\ref{compl}).

\section{\label{strategy} Fit strategy}
In the situation of an actual fit to values of $\P(Q^2)$ obtained from a numerical computation, these values are only known within some statistical errors.   That implies that we do not know any points of the function exactly.
However, the fact remains that $\P(Q^2)$ can be expressed in terms of a
Stieltjes function, \seef\ Eq.~(\ref{disprel}), and this implies that a series of
PA representations as described in Sec.~\ref{multipoint} exists that converges to
$\P(Q^2)$, when we parametrize the vacuum polarization as
\begin{equation}
\label{parvp}
\P(Q^2)=\P(0)-Q^2\left(a_0+\sum_{n=1}^N\frac{a_n}{b_n+Q^2}\right)\ .
\end{equation}
A number of PAs can be estimated by fitting this form to the data as a function
of increasing $N$.   For $a_0=0$ the parameters to be fitted are $\P(0)$ and
the $a_n$ and $b_n$ for $n\in\{1,\dots,N\}$, and we obtain an $[N-1,N]$ PA.
When also $a_0$ is fitted we obtain an $[N,N]$ PA, and together this sequence
of PAs estimates the standard PAs introduced in Sec.~\ref{multipoint}.  We will
enforce the  restrictions~(\ref{cond}) on the parameters $a_n$ and
$b_n$ in our fits (unless stated otherwise).  

Of course, since these PAs will not have been constructed as exact multi-point
PAs, the bounds described at the end of Sec.~\ref{multipoint} will not be exact
either.   However, we can still check the stability of our results as a function of
$N$.   In particular, we can check the stability of the value we obtain as a function of $N$ for the quantity $a_\m^{\rm HLO}$ defined in Eq.~(\ref{amu}).   As we will see, this quantity is much more stable as a function of $N$ than the
individual fit parameters in Eq.~(\ref{parvp}).   In practice, we have explored fits up to $N=3$; here we will present fits up to $N=2$.

As in Ref.~\cite{AB2007}, we will mostly explore fits in which we take the values of
$Q^2$ for which we fit the PAs from the data for $\P(Q^2)$ in an interval between
$Q^2=0$ and $Q^2=1$~GeV$^2$.   For each of our fits, we will compute the
quantity
\begin{equation}
\label{amuchopped}
a_\m^{{\rm HLO},Q^2\le 1}=4\a^2\int_0^{1\ {\rm GeV}^2} dQ^2\,f(Q^2)
\left(\P(0)-\P(Q^2)\right)\ ,
\end{equation}
with $f(Q^2)$ defined in Eq.~(\ref{amub}).   This of course misses the part of the 
integral between $1$~GeV$^2$ and $\infty$, but this part is of order a percent of the low-$Q^2$ contribution.   Since our goal here is to test the Pad\'e approach to fitting $\P(Q^2)$ for $Q^2\le 1$~GeV$^2$, we have
restricted ourselves to the expression in Eq.~(\ref{amuchopped}) for comparisons between different
fits.

We have not explored the complementary PAs defined in Sec.~\ref{multipoint}
yet, but we anticipate that they may become useful in the future, when more
precise data become available.

\section{\label{tests} Tests}
In this section, we explore fits to two different data sets.   One set is the data for $\P(Q^2)$ with light quark mass equal to $0.0124$ in lattice units
(on a $24^3\times 96$ lattice with lattice spacing $a\approx 0.09$~fm, using ``fine" configurations from the
MILC collaboration \cite{MILC}) that was also studied in Ref.~\cite{AB2007}
(see Table~I of that paper).    The other set is data obtained using
the MILC ``super-fine'' gauge configurations with lattice spacing $a\approx 0.06$~fm
on a $64^3\times 144$ lattice with light quark mass equal to $0.0018$
and a strange quark mass equal to $0.018$, in lattice units.   For both data
sets the lattice strange quark mass is approximately equal to the physical 
strange quark mass.
We will always
assume that the theory of Sec.~\ref{PA} applies to these data, \ie, that lattice
artifacts  are small enough to be ignored.
In practice,  rotational invariance is broken on the lattice.  Since $\P(Q^2)$ is extracted from
Eq.~(\ref{vacpol}) using a lattice definition of the momentum
components $Q_\m$ \cite{AB2007}, breaking of rotational invariance causes $\P(Q^2)$ to show small
deviations from the monotonic decrease that follows from
Eq.~(\ref{disprel}).\footnote{Also, all data have been obtained on ensembles of gauge configurations
generated with improved actions.   Therefore, the spectral function may not
be positive for values of the momenta near the lattice cutoff.} \\

\begin{table}[t]
\begin{center}
\begin{tabular}{|c|c|c|c|c|c|c|c|}
\hline
 & $\c^2$/dof & $10^{10}a_\m^{{\rm HLO},Q^2\le 1}$ & $\P(0)$ & $a_i$ & $b_i$ & $a_0$   \\
\hline
VMD & 5.86/3 & 363(7) & 0.0962(6) & 0.0471(9) & 0.9256(fixed) & -- \\
$[0,1]$ & 11.4/8 & 338(6) & 0.0960(5) & 0.0600(7) & 1.287(27) & -- \\
$[1,1]$ & 7.49/7 & 350(8) & 0.0963(6) & 0.049(4) & 1.09(9) & 0.0028(12) \\
$[1,2]$ & 7.49/6 & 350(8) & 0.0963(6) & 0.049(4) & 1.09(9) & -- \\
&&&& 2(17) & $2(8)\times 10^3$ & \\
$[2,2]$ & 7.49/5 & 350(7) & 0.0963(6) & 0.049(4) & 1.09(9) & 0.0012(10) \\
&&&& 2.4(1.4) & $1.4(0.8)\times 10^3$ & \\
\hline
\end{tabular}
\end{center}
\begin{quotation}
\floatcaption{a09cor}{{\it VMD and PA fits to the $a=0.09$~fm, $am_{light}=0.0124$ data for $\P(Q^2)$ of {\rm Ref.~\cite{AB2007}}
with $Q^2\le 0.6~{\rm GeV}^2$, except for the VMD fit, for which the
fit interval is $Q^2\le 0.35~{\rm GeV}^2$.  Correlated fits; $\c^2$ errors.   }}
\end{quotation}
\vspace*{-4ex}
\end{table}%

\begin{table}[t]
\begin{center}
\begin{tabular}{|c|c|c|c|c|c|c|c|}
\hline
 & $\c^2$/dof & $10^{10}a_\m^{{\rm HLO},Q^2\le 1}$ & $\P(0)$ & $a_i$ & $b_i$ & $a_0$   \\
\hline
VMD & 4.37/18 & 413(8) & 0.0980(7) & 0.0536(10) & 0.9256(fixed) & -- \\
$[0,1]$ & 3.58/17 & 373(37) & 0.0971(12) & 0.0569(25) & 1.10(16) & -- \\
$[1,1]$ & 3.36/16 & 424(116) & 0.0979(22) & 0.033(14) & 0.6(4) & 0.007(6) \\
$[1,2]$ & 3.35/15 & 443(293) & 0.098(4) & 0.02(10) & 0.4(1.7) & -- \\
&&&& 0.058(12) & 2(11) & \\
$[2,2]$ & 3.35/14 & 445(432) & 0.098(4) & 0.02(29) & 0.4(4.2) & 0.0(4) \\
&&&& 0.1(3.8) & 4(141) & \\
\hline
\end{tabular}
\end{center}
\begin{quotation}
\floatcaption{a09}{{\it VMD and PA fits to the $a=0.09$~fm, $am_{light}=0.0124$ data for $\P(Q^2)$ of {\rm Ref.~\cite{AB2007}}
with $Q^2\le 1~{\rm GeV}^2$.  Uncorrelated fits; errors computed by a linear
fluctuation analysis.   }}
\end{quotation}
\vspace*{-5ex}
\end{table}%

For each data set we carry out both correlated and uncorrelated fits, and compare those with each other.   It turns out that this raises interesting questions about the behavior of the data and the fits at very low $Q^2$.

\begin{boldmath}
\subsection{\label{fine} $a=0.09$~fm data at $m_{light}/m_{strange}=0.4$}
\end{boldmath}
For our first example we consider the $am_{light}=0.0124$ data that were also
considered in Ref.~\cite{AB2007}; this value of the light quark mass corresponds to about $2/5$ times the physical strange quark mass.  In Tables~\ref{a09cor} 
and \ref{a09} we show the result of a sequence of PA fits, with Table~\ref{a09cor} showing correlated fits, and Table~\ref{a09} showing uncorrelated fits.  For the correlated fits we fitted data on the interval $0<Q^2\le 0.6$~GeV$^2$, because this interval yields the smallest values for the $\c^2$ per degree of freedom.
For the uncorrelated case, we fitted the data for $\P(Q^2)$ on the
interval $0<Q^2\le 1$~GeV$^2$, as was done in Ref.~\cite{AB2007}.

\begin{figure}
\begin{center}
\includegraphics*[width=12cm]{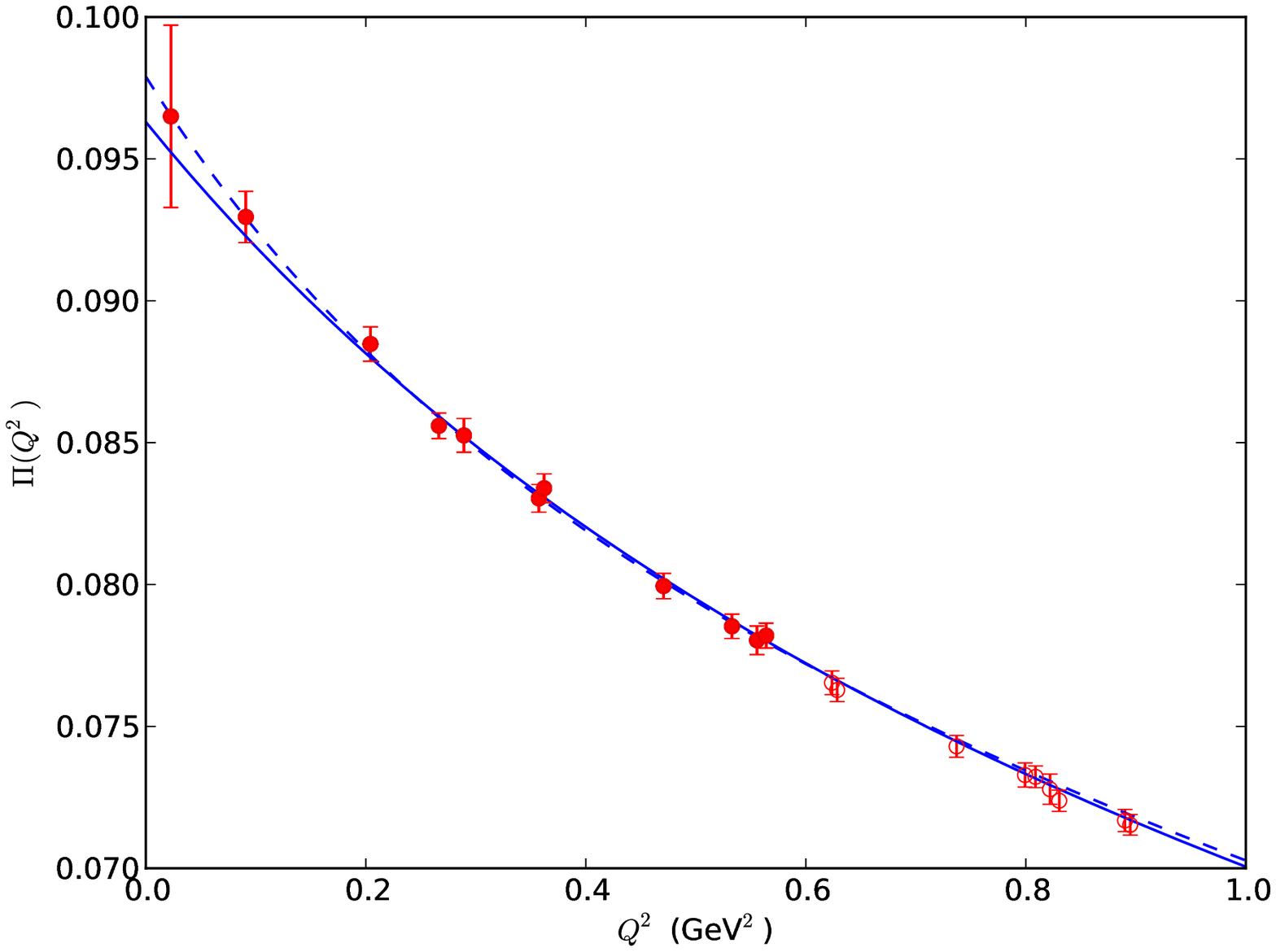}
\end{center}
\vspace*{-6ex}
\begin{quotation}
\floatcaption{table1and2line3}%
{{\it $[1,1]$  fits of Tables~\ref{a09cor} (correlated, solid curve) and \ref{a09} (uncorrelated, dashed curve) compared with data.
Solid points have been included in the correlated fit while both solid and open points have been included in the uncorrelated fit.}}
\end{quotation}
\vspace*{-4ex}
\end{figure}

 Table~\ref{a09cor} shows that the value for $a_\m^{{\rm HLO},Q^2\le 1}$ becomes very stable for PA fits starting at $[1,1]$.   For the $[1,1]$ PA 
 $\c^2/{\rm dof}=1.07$, indicating a good fit.   For higher PAs, the value of $\c^2$
 does not change, being very insensitive to the location of the second pole.
 It follows that the values of the parameters characterizing the second pole
 are not well determined, as can be seen in the table.   However, the value
 of $a_\m^{{\rm HLO},Q^2\le 1}$ is completely insensitive to the second pole.
The explanation for this is that the integral for $a_\m^{{\rm HLO},Q^2\le 1}$ is dominated by the $Q^2$ region around $m_\m^2$,
and thus very insensitive to the precise location of PA poles at large negative
values of $Q^2$.  
 
 The fit marked ``VMD'' is obtained by holding the parameter $b_1$ fixed
 at the square of the $\r$ mass (which is equal to 962~MeV for this data set) in what would otherwise be a 
 $[0,1]$ PA fit.   It is thus {\it not} one of the sequence of PAs introduced in
 Sec.~\ref{PA}.   According to the theory there is no reason one should expect 
 the parameter $b_1$ to be equal to the square of the $\r$ mass, as borne out by the values for $b_1$ found in the PA fits of
 Table~\ref{a09cor}.\footnote{See Sec.~\ref{discussion} for further discussion.}
    For the correlated VMD fit a fitting interval $0<Q^2\le 0.35$~GeV$^2$
    leads to the lowest $\c^2$ per degree of freedom.  With $\c^2/{\rm dof}\approx 2$, the VMD fit is not very good.   
   It is already much better for the $[0,1]$ PA,
 in which the constraint on $b_1$ is relaxed, and it decreases further, to an 
acceptable value, for the $[1,1]$ PA. 

Table~\ref{a09} shows similar fits, but here all fits are uncorrelated.
All errors have been estimated using a linear fluctuation analysis starting
from the uncorrelated $\c^2$, starting from the full data covariance matrix
\cite{BCGJMOP}.
These errors agree with errors computed under a single-elimination
jackknife.  In these PA fits we have
relaxed the constraint $b_1\ge 4m_\p^2=0.906$~GeV$^2$ (on this data set),
but one notes that the values of $b_1$ are consistent with this bound within
errors.   Both correlated and uncorrelated $[1,1]$
PA fits are shown in Fig.~\ref{table1and2line3}. 

The uncorrelated VMD fit reproduces ``fit A'' of Ref.~\cite{AB2007}, including the error.\footnote{The parameters
$\P(0)$ and $a_1$ are not the same as the parameters $A$ and $f_V$ of
Ref.~\cite{AB2007}.}   One notes that the uncorrelated PA fits lead to results consistent with those of Table~\ref{a09cor}, but with much larger errors.   The
uncorrelated VMD fit is not consistent with what we would expect to be the best fit,
\begin{equation}
\label{best09}
a_\m^{{\rm HLO},Q^2\le 1}=350(8)\times 10^{-10}\ ,
\end{equation}
from the $[1,1]$ PA of Table~\ref{a09cor}.

We may also compare the values in the tables with values obtained from
a fit with a fourth order polynomial in $Q^2$, which are
\begin{eqnarray}
\label{poly09s}
a_\m^{{\rm HLO},Q^2\le 1}&=&410(91)\times 10^{-10}\ ,\qquad\mbox{(uncorrelated)}\ ,\\
a_\m^{{\rm HLO},Q^2\le 1}&=&346(8)\times 10^{-10}\ ,\qquad\ \,\mbox{(correlated)}\ .\nonumber
\end{eqnarray}
The first line is in agreement with Ref.~\cite{AB2007}, and was fitted with $0<Q^2\le 1$~GeV$^2$,
as in Table~\ref{a09}, and the second is from a correlated fit on the 
interval $0<Q^2\le 0.6$~GeV$^2$, as in Table~\ref{a09cor}.  The latter fit has
a $\c^2$/dof of $7.48/6$, less good than the $[1,1]$ fit in Table~\ref{a09cor}. 
Both are in good agreement with Eq.~(\ref{best09}), given the size of the errors.

\vskip4ex
\begin{boldmath}
\subsection{\label{superfine} $a=0.06$~fm data at $m_{light}/m_{strange}=0.1$}
\end{boldmath}
For our second example, we consider the vacuum polarization computed on
MILC configurations at $a=0.06$~fm and $am_{light}=0.0018$, which is about
$1/10$ times the physical strange quark mass.   Correlated fits are
 shown  in Table~\ref{a06a}, where we fitted
the data for $0<Q^2\le 0.53$~GeV$^2$ (which corresponds to the 20 data points with the lowest values of $Q^2$).
The $\c^2$ values per degree of freedom of the fits in
Table~\ref{a06a} are slightly smaller than one, except for the VMD fit, for
which $\c^2$/dof is about two.\footnote{We thank Doug Toussaint for 
providing us with an unpublished rough estimate of the $\r$ mass for
this data set.}
  We find that the value of $\c^2/$dof
increases if we fit over a larger range of $Q^2$ values, and we will therefore
take the results of Table~\ref{a06a} as our optimal results (for more
on this point, see the discussion around Table~\ref{padevspoly} below).
Uncorrelated fits are shown in
Table~\ref{a06auncor}, where, in line with Sec.~\ref{fine}, fits were carried out on the interval $0<Q^2\le 1$~GeV$^2$.   

\begin{table}[t]
\begin{center}
\begin{tabular}{|c|c|c|c|c|c|c|c|}
\hline
 & $\c^2$/dof & $10^{10}a_\m^{{\rm HLO},Q^2\le 1}$ & $\P(0)$ & $a_i$ & $b_i$ & $a_0$   \\
\hline
VMD & 38.6/18 & 646(8) & 0.1222(6) & 0.0595(8) & 0.64 (fixed) & -- \\
$[0,1]$ & 14.3/17 & 550(20) & 0.1203(7) & 0.0646(16) & 0.83(5) & -- \\
$[1,1]$ & 13.9/16 & 572(41) & 0.1206(8) & 0.052(16) & 0.68(20) & 0.005(7) \\
$[1,2]$ & 13.9/15 & 572(37) & 0.1206(8) & 0.052(14) & 0.68(19) & -- \\
&&&& 1(6) & $0.3(1.0)\times 10^3$ & \\
$[2,2]$ & 13.9/14 & 572(38) & 0.1206(8) & 0.052(14) & 0.68(18) & 0.003(27) \\
&&&& 1(31) & $0.4(6.0)\times 10^3$ & \\
\hline
\end{tabular}
\end{center}
\begin{quotation}
\floatcaption{a06a}{{\it PA fits to the $a=0.06$~fm, $am_{light}=0.0018$ data  for $\P(Q^2)$
with $Q^2\le 0.53~{\rm GeV}^2$.  Correlated fits; $\c^2$ errors.
 }}
\end{quotation}
\vspace*{-4ex}
\end{table}%

\begin{table}[t]
\begin{center}
\begin{tabular}{|c|c|c|c|c|c|c|c|}
\hline
 & $\c^2$/dof & $10^{10}a_\m^{{\rm HLO},Q^2\le 1}$ & $\P(0)$ & $a_i$ & $b_i$ & $a_0$   \\
\hline
VMD & 37.2/51 & 685.2(7.8) & 0.1236(6) & 0.0631(7) & 0.64 (fixed) & -- \\
$[0,1]$ & 13.9/50 & 555(22) & 0.1208(8) & 0.0666(7) & 0.85(4) & -- \\
$[1,1]$ & 12.0/49 & 645(66) & 0.1221(13) & 0.047(5) & 0.54(11) & 0.0071(21) \\
$[1,2]$ & 11.4/48 & 788(482) & 0.123(4) & 0.015(20) & 0.2(4) & -- \\
&&&& 0.063(14) & 1.4(9) & \\
$[2,2]$ & 11.3/47 & 837(627) & 0.124(5) & 0.018(5) & 0.2(5) & 0.022(9) \\
&&&& 0.22(6) & 3.9(6) & \\
\hline
\end{tabular}
\end{center}
\begin{quotation}
\floatcaption{a06auncor}{{\it PA fits to the $a=0.06$~fm, $am_{light}=0.0018$ data for $\P(Q^2)$
with $Q^2\le 1~{\rm GeV}^2$.  Uncorrelated fits;  errors from linear fluctuation
analysis.   For the $[1,2]$ and $[2,2]$ fits, $b_1$ is at the limit $4m_\p^2=0.1936$~GeV$^2$ (for this ensemble),
which was enforced in those fits.
 }}
\end{quotation}
\vspace*{-4ex}
\end{table}%

\begin{figure}[t]
\begin{center}
\includegraphics*[width=12cm]{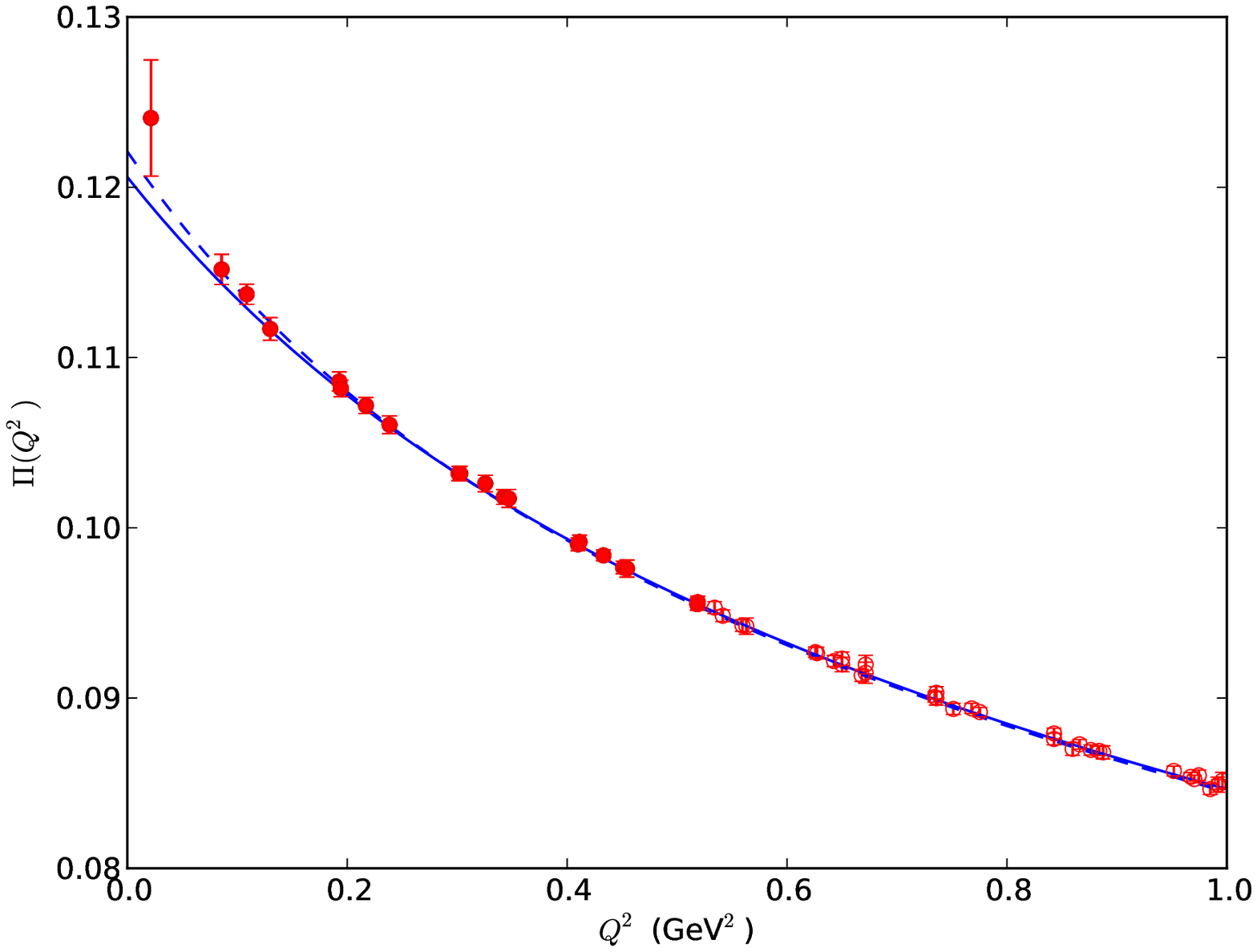}
\end{center}
\vspace*{-6ex}
\begin{quotation}
\floatcaption{table3and4line3}%
{{\it $[1,1]$ PA fits of Tables~\ref{a06a} (correlated, solid curve) and \ref{a06auncor} (uncorrelated, dashed curve) compared with data.
Solid points have been included in the correlated fit while both solid and open points have been included in the uncorrelated fit.}}
\end{quotation}
\vspace*{-4ex}
\end{figure}

\begin{figure}[ht]
\begin{center}
\includegraphics*[width=12cm]{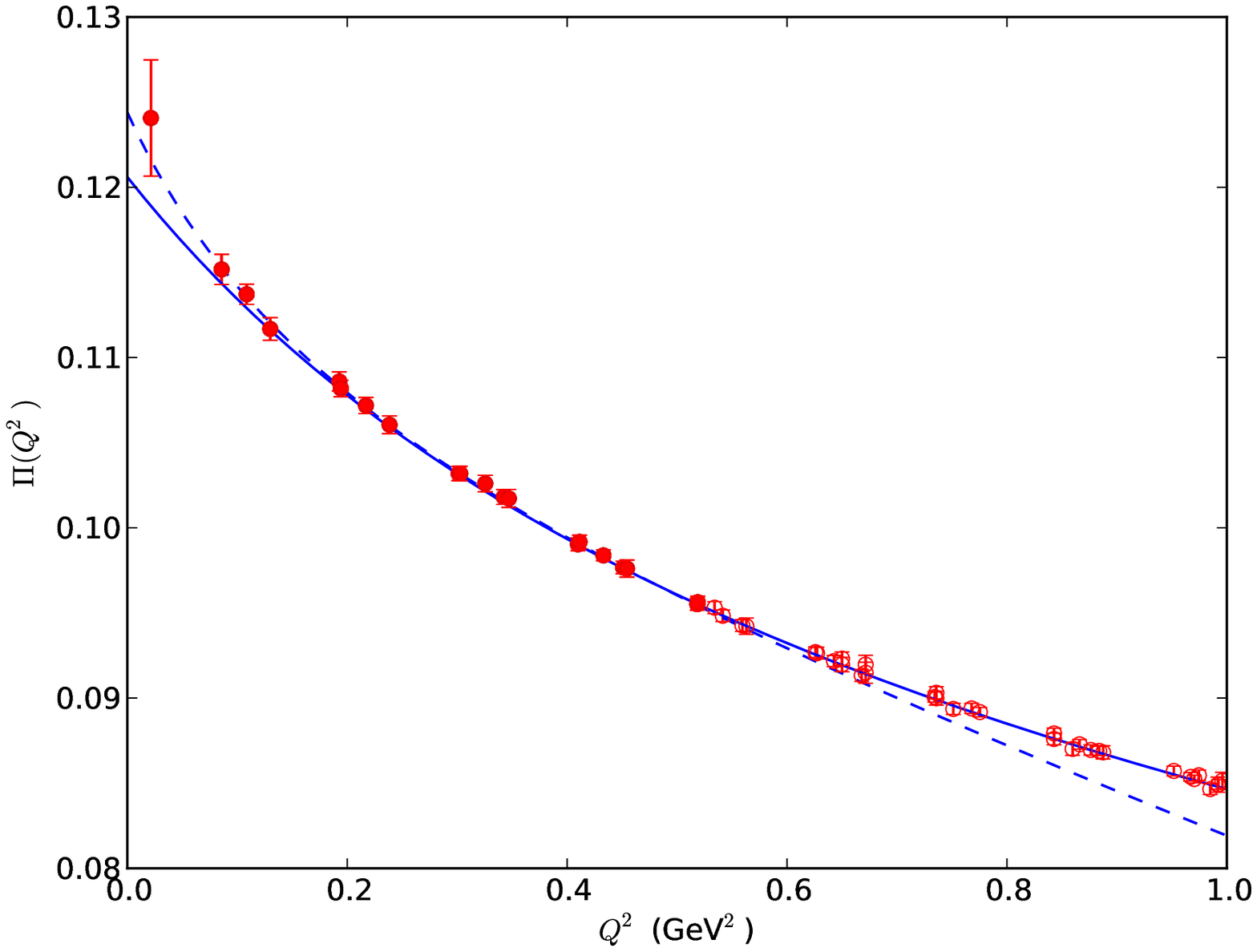}
\end{center}
\vspace*{-6ex}
\begin{quotation}
\floatcaption{table3line3andunc}%
{{\it $[1,1]$ correlated (solid curve) and uncorrelated (dashed curve)
fits as in Table~\ref{a06a}, fitted on interval
$0<Q^2\le 0.53$~{\rm GeV}$^2$.  
Solid points have been included in the fits, open points have not been included.}}
\end{quotation}
\vspace*{-4ex}
\end{figure}

It is again not surprising that the correlated fits become less good if one fits over a larger
range in $Q^2$.   As before, it is clear from the tables that, given the
quality of the data, it is very hard to fit a second pole.   The value of
$a_\m^{{\rm HLO},Q^2\le 1}$ is again completely insensitive to the location of
the second pole.\footnote{We even considered $[2,3]$ and $[3,3]$ fits, with
the conclusion being the same.}   

We show the $[1,1]$ fits of Tables~\ref{a06a} and \ref{a06auncor} in Fig.~\ref{table3and4line3}.  As in Fig.~\ref{table1and2line3}
one notes the sensitivity of the fit near $Q^2=0$; this
explains the different values for $a_\m^{{\rm HLO},Q^2\le 1}$ shown in the
tables.

{}From the $[1,1]$ PA fit of Table~\ref{a06a} we take what we would
expect to be our best result for this data set:
\begin{equation}
\label{06result}
a_\m^{{\rm HLO},Q^2\le 1}=572(41)\times 10^{-10}\ .
\end{equation}

In Fig.~\ref{table3line3andunc} we show correlated and  uncorrelated $[1,1]$ PA fits,
now taking the range $0<Q^2\le 0.53$~GeV$^2$ as our fitting range also
for the uncorrelated fit.
We note that the uncorrelated fit appears to do better than the uncorrelated $[1,1]$
PA fit shown in Fig.~\ref{table3and4line3} at the lowest $Q^2$ value,
but much less well than the correlated fit for $Q^2>0.53$~GeV$^2$.
Accordingly, uncorrelated fits are quite sensitive to the fitting range.
For instance, the central value of $a_\m^{{\rm HLO},Q^2\le 1}$ from the 
uncorrelated fit shown in Fig.~\ref{table3line3andunc} is 42\% larger than
from a similar fit on the range $0<Q^2\le 1$~GeV$^2$ (shown in Table~\ref{a06auncor}).   A correlated fit on the latter
range gives a central value which is only 3\% larger than the value in Eq.~(\ref{06result}),
\ie, it is within the error given in that equation.\footnote{Despite the fact that for a correlated fit on the range $0<Q^2\le 1$~GeV$^2$ the value of
$\c^2$ is about 2.5 per degree of freedom.}

In Ref.~\cite{AB2007} also polynomial fits with third- and fourth-order polynomials
were considered, and it is thus
interesting to compare PA fits with polynomial fits.   For the data of this
subsection, the radius of convergence, $4m_\p^2=0.194$~GeV$^2$.\footnote{The radius of convergence for the case of Sec.~\ref{fine} is
much larger, which is why we chose to make this comparison in this
subsection.}
Therefore, fitted polynomials cannot be interpreted as estimates of the
Taylor expansion of $\P(Q^2)$ around $Q^2=0$, as long as we use a
fitting interval with upper bound larger than $4m_\p^2$.

We show third- and fourth-order polynomial fits, as well as $[1,1]$ and
$[1,2]$ PA fits in Table~\ref{padevspoly}, as a function of the number
of data points in the fit (20 points corresponds to the fitting interval
$0<Q^2\le 0.53$~GeV$^2$ used in Table~\ref{a06a}).  All fits shown
are correlated.   Both ``Poly~3'' and ``PA~[1,1]'' are four-parameter fits,
while ``Poly~4'' and ``PA~[1,2]'' are five-parameter fits.  The $\c^2/$dof
for all fits is good, except for fits with 26 data points, for which it shows
a steep increase.   

We observe that Poly~3, PA~[1,1] and PA~[1,2] fits all lead to values for
$a_\m^{{\rm HLO},Q^2\le 1}$ which are stable within the error given in
Eq.~(\ref{06result}).  For the Poly~4 fit, however, this spread is much larger.
Adding a fit parameter by going from Poly~3 to
Poly~4 fits leads to significant changes in the central value for
$a_\m^{{\rm HLO},Q^2\le 1}$, whereas  going from
PA~[1,1] to PA~[1,2] fits the central values do not change much.   In other words,
if we would do a correlated Poly~4 fit to 20 data points, for which we
would find $a_\m^{{\rm HLO},Q^2\le 1}=535(45)\times 10^{-10}$, the 
error would be underestimated because of the spread of values for the
Poly~4 fit shown in Table~\ref{padevspoly}, while the error shown in
Eq.~(\ref{06result}) encompasses the full range of $a_\m^{{\rm HLO},Q^2\le 1}$
PA results shown in Table~\ref{padevspoly}.   The Poly~4 and PA~[1,2] fits
with 20 data points are shown in Fig.~\ref{table3line4andpoly}.

\begin{table}[t]
\begin{center}
\begin{tabular}{|c|c|c|c|c|c|c|c|c|}
\hline
& \multicolumn{2}{c|}{Poly~3} & \multicolumn{2}{c|}{Poly~4}
& \multicolumn{2}{c|}{PA~[1,1]} & \multicolumn{2}{c|}{PA~[1,2]}\\
\hline\hline
\# points & $\c^2$/dof & $a_\m^{(1)}$
& $\c^2$/dof & $a_\m^{(1)}$
& $\c^2$/dof & $a_\m^{(1)}$
& $\c^2$/dof & $a_\m^{(1)}$\\
\hline
16 & 9.6/12 & 543 &  9.5/11 & 483 & 9.7/12 & 564 & 9.7/11 & 565  \\
18 & 11.4/14 & 526 & 10.5/13 & 596 & 11.2/14 & 541 & 11.5/13 & 561   \\
20 & 13.1/16 & 536 & 13.1/15 & 535 & 13.9/16 & 572 & 13.9/15 & 572  \\
22 & 16.5/18 & 541 & 15.9/17 & 513 & 18.5/18 & 566 & 18.5/17 & 566 \\
24 & 16.6/20 & 537 & 16.4/19 & 521 & 19.4/20 & 583 & 19.4/19 & 583 \\
26 & 30.7/22 & 505 & 23.6/21 & 580 & 26.8/22 & 557 & 26.7/21 & 560  \\
\hline
\end{tabular}
\end{center}
\begin{quotation}
\floatcaption{padevspoly}{{\it Correlated PA and polynomial fits to the
data of Table~\ref{a06a}, as a function of the fitting interval.
The first column shows the number of data points in the fit, with
20 points corresponding to the fitting interval $0<Q^2\le 0.53~{\rm GeV}^2$
of Table~\ref{a06a}.
   The column {\rm ``Poly}~$n${\rm ''}  shows
results from a fit to a polynomial of degree $n$;
$a_\m^{(1)}$ stands for $10^{10}a_\m^{{\rm HLO},Q^2\le 1}$.  }}
\end{quotation}
\vspace*{-4ex}
\end{table}%

\begin{figure}[t]
\begin{center}
\includegraphics*[width=12cm]{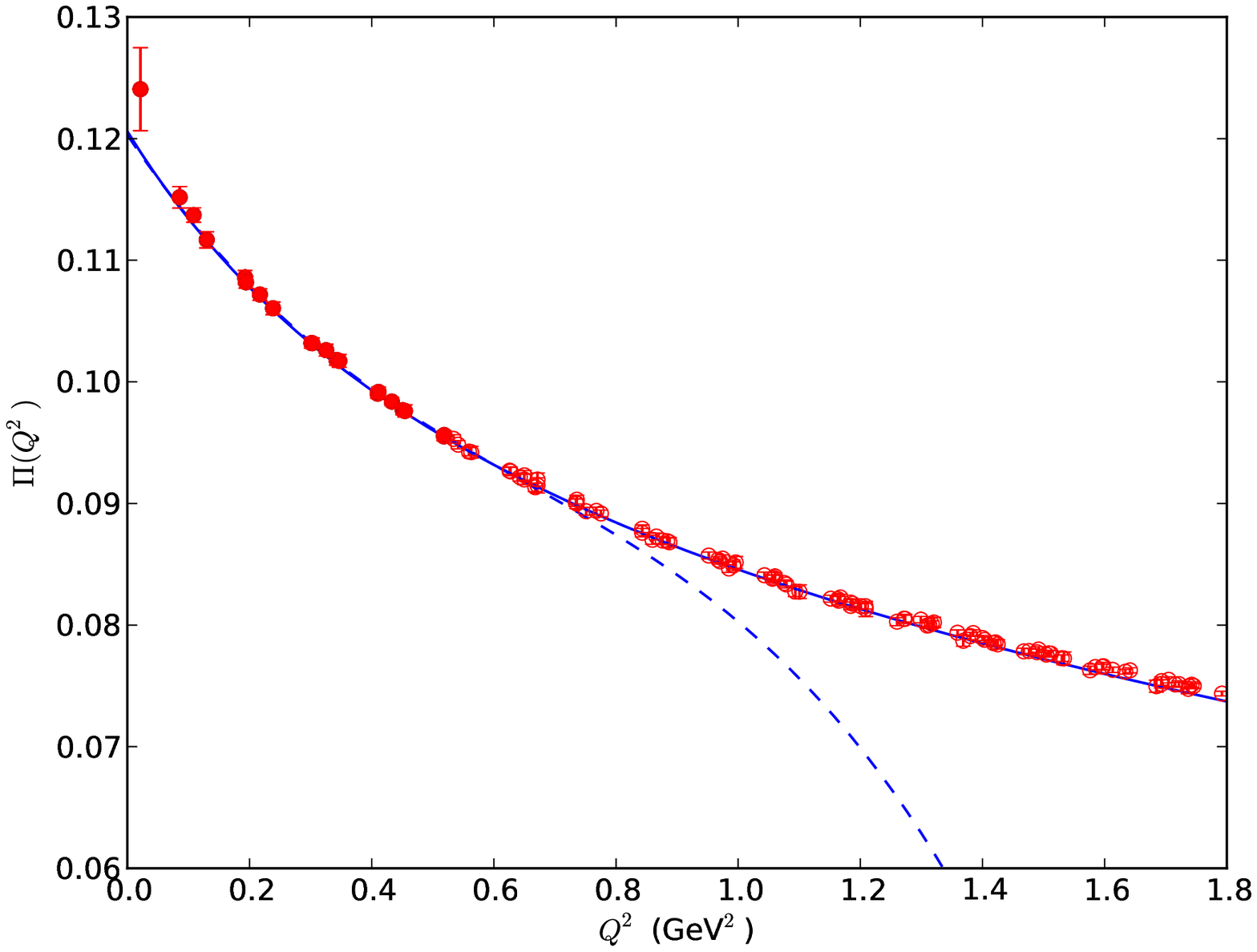}
\end{center}
\vspace*{-6ex}
\begin{quotation}
\floatcaption{table3line4andpoly}%
{{\it Comparison of correlated $[1,2]$ PA (solid curve) and 4th-order polynomial (dashed curve) fits, both fitted on the interval
$0<Q^2\le 0.53$~{\rm GeV}$^2$.
Solid points have been included in the fits, open points have not been included.}}
\end{quotation}
\vspace*{-4ex}
\end{figure}

\vskip4ex
\subsection{\label{discussion} Discussion of fits}
In this subsection we will discuss the fit results presented in Tables~\ref{a09cor} through \ref{a06auncor} in more detail.   We begin with the
$a=0.09$~fm results of Tables~\ref{a09cor} and \ref{a09}.

It is important to emphasize again that the VMD fits are not part of the
sequence of PAs introduced in Sec.~\ref{PA}, because in the VMD fits the
pole at $Q^2=-b_1$ is held fixed at the estimated (squared) $\r$ mass on this
ensemble.   The actual QCD spectral function has a cut on the negative
axis starting at $Q^2=-4m_\p^2$; any poles reside on the second Riemann
sheet, away from the negative axis.\footnote{Only in the limit of an
infinite number of colors do the poles move toward the negative real axis, and the
vacuum polarization becomes a meromorphic function.}   While positivity of the spectral
function implies that $\P(Q^2)$ can be expressed in terms of a Stieltjes
function, with a convergent sequence of PAs given by Eq.~(\ref{standpar}),
there is no reason that any of the poles of these Pad\'e's should
be equal to (the real part of) any pole representing a resonance in QCD.
In particular, in the PA fits, the parameter $b_1$ should not be taken
equal to the square of the $\r$ mass, but instead it should be left as a free parameter.
We included the VMD fits in Tables~\ref{a09cor} and \ref{a09} in order
to compare them with the PA fits.

First, we note that the correlated VMD fit in Table~\ref{a09cor} is a rather
poor fit, with a high $\c^2$/dof, and there is no agreement between the
correlated and uncorrelated VMD fits.
The quality of the correlated fits improves when
we add more parameters, first by varying $b_1$ and then by adding in the
parameter $a_0$, by which time $\c^2/\mbox{dof}\approx 1$.

The next observation is that the correlated fits do not get better by adding
a second pole to the PA.   The minimum value of $\c^2$ stays the same, and consequently,
the parameters of the second pole are very poorly determined.   We have
checked
that this does not depend on the fitting range employed.   A possible
explanation is that the lattice data for the vacuum polarization do not
quite follow the behavior predicted by Eq.~(\ref{disprel}), because of the
breaking of rotational invariance on the lattice (\seef\ discussion at the 
beginning of this section).   This can be seen from
the fact that the data points in Fig.~\ref{table1and2line3} show small
deviations from a smooth, monotonically-decreasing behavior.   However,
we note that the value of $a_\m^{{\rm HLO},Q^2\le 1}$ is completely
insensitive to the parameters of the second (and higher) poles.   Our best
correlated fit value for $a_\m^{{\rm HLO},Q^2\le 1}$ is given in 
Eq.~(\ref{best09}).

\begin{figure}[t]
\begin{center}
\includegraphics*[width=12cm]{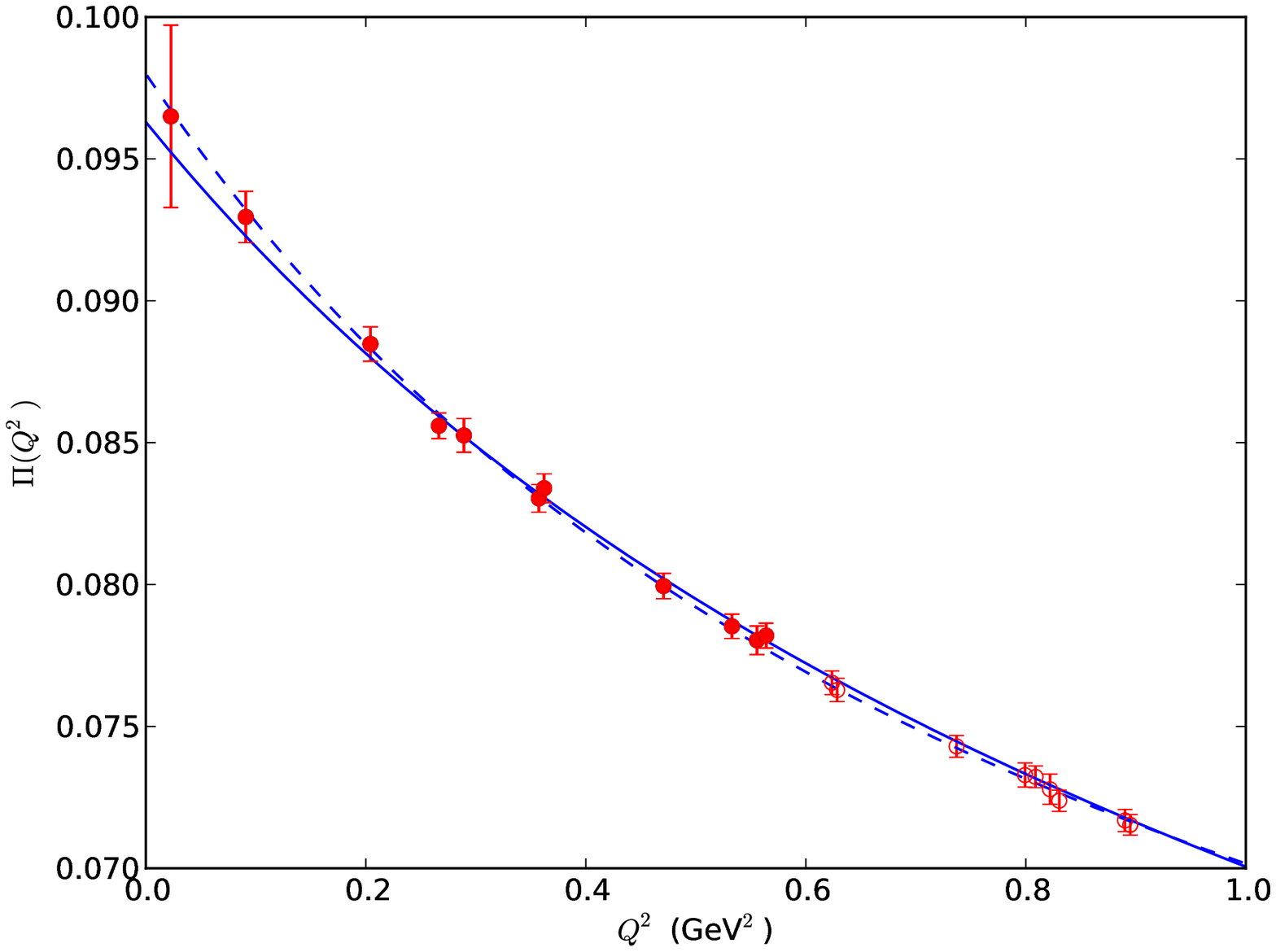}
\end{center}
\vspace*{-6ex}
\begin{quotation}
\floatcaption{table1line3table2line1}%
{{\it $[1,1]$  fit of Table~\ref{a09cor} (correlated, solid curve) and
VMD fit of Table~\ref{a09} (uncorrelated, dashed curve) compared with data.
Solid points have been included in the correlated fit while both solid and open points have been included in the uncorrelated fit.}}
\end{quotation}
\vspace*{-4ex}
\end{figure}

Moving to the uncorrelated fits of Table~\ref{a09}, we observe that the values
for $a_\m^{{\rm HLO},Q^2\le 1}$ for all fits in that table are consistent
with each other, because of the rapid increase of errors with the order of
the PA.   Furthermore, all uncorrelated and correlated PA fits are consistent
with each other as well, but clearly the correlated fits have much smaller
errors.

It is instructive to compare the best correlated fit, the $[1,1]$ PA fit, with the
uncorrelated VMD fit, because both have very small errors, and provide
a good fit, as can be seen in Fig.~\ref{table1line3table2line1}.   Both fits
are good fits, but they lead to values for $a_\m^{{\rm HLO},Q^2\le 1}$
which are not consistent with each other.   The statistical error on the uncorrelated VMD fit is very small, but this fit has an unknown systematic
error because of its model dependence.   This may explain the discrepancy
with the correlated $[1,1]$ PA fit.   The latter may be expected to have a
much smaller systematic error, since it is known to be a member of a 
converging sequence of PAs, and clearly already provides a good fit also
in the larger $Q^2$ region that was not included in the fit (the horizontal
axis of Fig.~\ref{table1line3table2line1} covers about twice the fitted region $0<Q^2\le 0.53$~GeV$^2$).   

However, for the computation of $a_\m^{{\rm HLO},Q^2\le 1}$ the region
$Q^2\sim m_\m^2=0.011$~GeV$^2$ dominates, and it is clear that the
data do not distinguish between these two fits in that region.   While the
rapid increase of the goodness of fit seen in Table~\ref{a09cor} from the
VMD fit to the $[1,1]$ fit can be taken as an indication that correlated
fits unbiased by model dependence are promising, we conclude that it
is not possible to exclude either value of $a_\m^{{\rm HLO},Q^2\le 1}$
on the basis of these data.

Similar remarks apply to the fits shown in Tables~\ref{a06a} and \ref{a06auncor}.   Correlated PA fits all have $\c^2/{\rm dof}\approx 1$,
unlike the correlated VMD fit for which $\c^2/{\rm dof}\approx 2$.
In both cases, the correlated and uncorrelated VMD fits do not agree
within errors (which, we recall, are purely statistical).   The 
value we obtain for $a_\m^{{\rm HLO},Q^2\le 1}$ is larger than that
obtained in Sec.~\ref{fine}; we believe that this is mostly due to a
smaller pion mass, with $m_\p\approx 220$~MeV for this data set,
while $m_\p\approx 480$~MeV for the $a=0.09$~fm data set.

\section{\label{conclusion} Conclusion}
In this article, we presented a new, model-independent method for
fitting the hadronic vacuum polarization $\P(Q^2)$ as a function of euclidean
momentum $Q^2$ to data obtained from a lattice QCD computation.   The
method is based on the theory of Pad\'e approximants (PAs) to a Stieltjes
function, and yields, in principle, a converging sequence of PAs to the vacuum polarization.

These PAs can be used to obtain lattice estimates
for the leading hadronic contribution to the anomalous magnetic moment
of the muon from Eq.~(\ref{amu}).   By comparing successive PAs in the sequence, one should be able to check the convergence in practice.
This would allow for a fully model-independent determination of the leading
hadronic contribution $a_\m^{\rm HLO}$, and thus help eliminate an
unknown systematic error present in all lattice
computations of $a_\m^{\rm HLO}$ to date.

In comparison with the VMD fits which have been employed in the literature,
these PAs contain more parameters (the $[0,1]$ PA already contains
three parameters, whereas the simplest VMD {\it ansatz} contains only two).
One thus typically expects larger statistical errors given certain lattice data.
However, the PA approach avoids model-dependent assumptions,
and hence removes the unknown systematic error associated with the
VMD approach.   

We have explored this new framework on two state-of-the-art ensembles
of gauge configurations, at different lattice spacings and pion masses.
{}From these explorations, we conclude that this new method looks
promising, but that better data at very low values of $Q^2$ will be 
needed in order to control the extrapolation necessary for a reliable 
computation of $a_\m^{\rm HLO}$ from the integral in Eq.~(\ref{amua}).  
 
Our explorations show that given current lattice data for $\P(Q^2)$,
there is a significant difference between our best PA fits (which are
four-parameter $[1,1]$ PAs), and VMD fits.\footnote{We recall that VMD fits
cannot be viewed as low-order PA fits, because there is no {\it a priori}
relation between PA poles and QCD resonance parameters; 
consequently, the
first pole should not be chosen equal to the square of the $\r$ mass,
as is done in most VMD fits.}  For instance, the difference between the
values of $a_\m^{{\rm HLO},Q^2\le 1}$ obtained from
the correlated $[1,1]$ PA fit of Table~\ref{a09cor} and the uncorrelated
VMD fit of Table~\ref{a09} and Ref.~\cite{AB2007} is about 15-20\%, much
larger than the statistical fit error on each of these values.   While it is
tempting to view the value from the correlated $[1,1]$ PA fit as the more
reliable one, it is clear from Fig.~\ref{table1line3table2line1} that
more data points with higher precision at low $Q^2$ are needed in order
to reduce this uncertainty.

Our explorations also showed that with these data it is very difficult to
fit the parameters characterizing the second and higher poles of the
PAs.   In order to test the convergence of the sequence of PAs fitted 
to $\P(Q^2)$, it would be desirable to investigate this issue, which
is possibly related to breaking of rotational invariance at non-zero
lattice spacing, in more detail in the future.   This issue appears to 
have no direct effect on the value of $a_\m^{{\rm HLO},Q^2\le 1}$,
which we found to be very insensitive to the location and residues of the second and higher poles.

In conclusion, the new method presented here looks promising, but 
data for $\P(Q^2)$ with more values at $Q^2\sim m_\m^2$ and with
higher statistics will be necessary in order to attain the high precision
determination of $a_\m^{\rm HLO}$ needed for a meaningful 
comparison with experiment.   Work in this direction is in progress.

\vspace{3ex}
\noindent {\bf Acknowledgments}
\vspace{3ex}

We would like to thank USQCD for the computing resources used to generate the vacuum polarization as well as the MILC collaboration for providing the configurations used.
TB and MG are supported in part by the US Department of Energy under
Grant No. DE-FG02-92ER40716 and Grant No. DE-FG03-92ER40711.
SP is supported by CICYTFEDER-FPA2008-01430, FPA2011-25948, SGR2009-894,
the Spanish Consolider-Ingenio 2010 Program
CPAN (CSD2007-00042) and also by the Programa de Movilidad
PR2010-0284.


\end{document}